\documentclass[a4paper, 11pt, oneside]{article}
\pdfoutput=1
\usepackage[a4paper, top=30mm, bottom=30mm, left=30mm, right=30mm]{geometry}
\usepackage{setspace} 
\usepackage[utf8]{inputenc}
\usepackage[T1]{fontenc}
\usepackage{lmodern} 
\usepackage[english]{babel}
\usepackage[hyphens]{url}
\usepackage[pdftex,  
            , pdfauthor={Pietro Coretto and Christian Hennig},
            , pdftitle={Robust improper maximum likelihood:  
                        tuning, computation, and a comparison with  other 
                        methods for robust Gaussian clustering},
            , pdfsubject={Statistics},
            , pdfkeywords={Cluster analysis, EM-algorithm, improper density, 
                           maximum likelihood, mixture models, model-based clustering, robustness},
            , pdfproducer={},
            , pdfcreator={PdfTeX; Emacs; Linux OS}
            , colorlinks=true
            , citecolor=blue
            , linkcolor=blue
            , urlcolor=blue
            , breaklinks=true
            , pdftex=true]{hyperref}
\usepackage{etoolbox}
\usepackage{amsmath, amsthm, amssymb, amsfonts}
\usepackage{color}
\usepackage{natbib} 
\usepackage{graphicx}
\usepackage{subfigure}
\usepackage{caption}
\usepackage{amsmath, amsthm, amsfonts, amssymb}
\usepackage{float}
\usepackage{booktabs}
\usepackage{datetime}

\newcommand{\set}[1]{\left\{ #1 \right\}}%
\newcommand{\abs}[1]{\left| #1 \right| }%
\newcommand{\ind}{\text{\bf 1}}
\def\argmin{\mathop{\rm arg\,min}\limits}
\def\argmax{\mathop{\rm arg\,max}\limits}
\newcommand{\R}{\mathbb{R}}
\newcommand{\N}{\mathbb{N}}

\newcommand{\E}{{\rm E}}
\newcommand{\pimax}{\pi_{\max}}
\newcommand{\lmax}{\lambda_{\max}}
\newcommand{\lmin}{\lambda_{\min}}
\newcommand{\js}{{j=1,2,\ldots,G}}
\newcommand{\ks}{{k=1,2,\ldots,p}}

\theoremstyle{plain}

\theoremstyle{definition}
\newtheorem{definition}{Definition}

\numberwithin{equation}{section}

\newcounter{Subitem}
\begin{document}

\date{}

\title{\Large  ROBUST IMPROPER MAXIMUM LIKELIHOOD: \\  TUNING, COMPUTATION, AND A COMPARISON WITH \\ OTHER   METHODS FOR ROBUST GAUSSIAN CLUSTERING}
\author{{Pietro Coretto%
\footnote{The author gratefully acknowledges support from the University of Salerno grant program ``Sistema di calcolo ad alte prestazioni per l'analisi economica, finanziaria e statistica (HPC - High Performance Computing) - prot. ASSA098434, 2009''}} \\ University of Salerno, Italy \\ {e-mail: pcoretto@unisa.it} 
\and %
{Christian Hennig\footnote{The author gratefully acknowledges support from the  EPSRC grant EP/K033972/1. }} \\ {University College London, UK} \\ {e-mail: c.hennig@ucl.ac.uk} }

\thispagestyle{empty}
\maketitle

{\color{blue} \textbf{{\centering \sf This is a preprint. The revised version of this paper is published as\\}}}
\vspace{.5em}
Coretto, P. and Hennig, C. (2016).  Robust Improper Maximum Likelihood: Tuning, Computation, and a Comparison with Other Methods for Robust Gaussian Clustering. \textit{Journal of the American Statistical Association} 111(516), pp. 1648--1659. $[$\href{http://dx.doi.org/10.1080/01621459.2015.1100996}{\sf link}$]$

\vspace{1em}

\begin{quote}{
\small {\bf Abstract.}~The two main topics of this paper are the introduction of the ``optimally tuned improper maximum likelihood estimator'' (OTRIMLE) for robust clustering based on the multivariate Gaussian model for clusters, and a comprehensive simulation study comparing the OTRIMLE to Maximum Likelihood in Gaussian mixtures with and without noise component, mixtures of t-distributions, and the TCLUST approach for trimmed clustering. The OTRIMLE uses an improper constant density for modelling outliers and noise. This can be chosen optimally so that the non-noise part of the data looks as close to a Gaussian mixture as possible. Some deviation from Gaussianity can be traded in for lowering the estimated noise proportion. Covariance matrix constraints and computation of the OTRIMLE are also treated. In the simulation study, all methods are confronted with setups in which their model assumptions are not exactly fulfilled, and in order to evaluate the experiments in a standardized way by misclassification rates, a new model-based definition of ``true clusters'' is introduced that deviates from the usual identification of mixture components with clusters. In the study, every method turns out to be superior for one or more setups, but the OTRIMLE achieves the most satisfactory overall performance. The methods are also applied to two real datasets, one without and one with known ``true'' clusters.\\


}
\end{quote}


\section{Introduction}\label{sec_introduction}
In this paper we introduce and investigate the 
``optimally tuned robust improper maximum likelihood estimator''
(OTRIMLE), a method for robust 
clustering with clusters that can be
approximated by multivariate Gaussian distributions. Its one-dimensional
version was
introduced in \cite{Coretto_Hennig_2010}. 
We also present a simulation study comparing OTRIMLE and other
approaches for (mostly robust) model-based clustering, 
which is, to our knowledge,
the most comprehensive study in the field and involves a careful discussion of
the issue of comparing methods based on different model assumptions. 

The basic idea of OTRIMLE is to fit an improper density to the data that
is made up by a Gaussian mixture density and a ``pseudo mixture component'' 
defined by a small constant density, which is meant to capture outliers in low
density areas of the data. This is inspired by the addition of
a uniform ``noise component'' to a Gaussian mixture
(\cite{Banfield_Raftery_1993}). \cite{Hennig_2004} showed
that using an improper density improves the breakdown robustness of this
approach.  The OTRIMLE has been found to work well for 
one-dimensional data in 
\cite{Coretto_Hennig_2010}.   

As in many other statistical problems, violations of the model assumptions
and particularly outliers may cause problems in cluster analysis. Our general
attitude to the use of statistical models in cluster analysis is that the 
models should not be understood as reflecting some underlying but in practice
unobservable ``truth'', but rather as thought constructs implying a certain
behaviour of methods derived from them (e.g., maximizing the likelihood),
which may or may not be appropriate in a given application (more details
on the general philosophy of clustering can be found in 
\cite{Hennig_Liao_2013}). Using a model such as a mixture of multivariate
Gaussian distributions, interpreting every mixture component as a ``cluster'',
implies that we look for clusters that are approximately ``Gaussian-shaped'',
but we do not want to rely on whether the data really were generated i.i.d. by 
a Gaussian mixture. We are interested in
the performance of such methods in situations where one may legitimately look
for Gaussian-shaped clusters, even if some data points do not belong to such 
clusters (called ``noise'' in the following), 
and even if the clusters are not precisely Gaussian. This 
reflects the fact that in practice for example mixtures of $t$-distributions
are used for clustering the same data sets to which Gaussian mixtures
are fitted as well, interpreting the resulting clusters in the same way.

For illustration of the outlier problem in model-based clustering, we use
a 5-dimensional 
data set in which the 170 districts of the German city of Dortmund are
characterized by a number of variables, which is discussed in detail in Section 
\ref{sec_dortmund}. Fitting a plain Gaussian mixture with $G=4$ to all five 
variables by R's MCLUST package (\cite{Mclust_Software_Manual_2006}), one 
cluster is a one-point cluster consisting only of an extreme outlier, and two
further clusters fit two different varieties of moderate outliers. More than
120 districts are collected in a single cluster. The task of robust clustering
is to avoid having many or even most clusters dominated by outliers, and to
produce a meaningful clustering structure also among the main bulk of 
non-extreme observations.

A number of model-based clustering methods that can deal with outliers have  been proposed in recent years. An overview of these methods is given in Section \ref{sec_methods_from_literature}.   The OTRIMLE is introduced and discussed  in Section \ref{sec_otrimle}, starting from the ``RIMLE'', in which the level of the improper constant density is a tuning constant.  We then introduce a method for optimal tuning and discuss its computation. 
Section \ref{sec_simulation_study} presents a simulation study that uses a unified approach for defining elliptically shaped clusters with noise/outliers, presented in Section \ref{sec_true_clusters}. The Dortmund data set mentioned above is discussed in Section \ref{sec_application} along with a dataset of folk song melodies from two known regions. Some further issues including the estimation of the number of clusters are discussed in Section \ref{sec_concluding_remarks}. Additional details about the example dataset (including full scatterplots), the simulation study, and computation of methods are provided in an online supplement (\cite{Coretto_Hennig_2014_supplement}). Theoretical properties of the RIMLE with the tuning constant fixed 
are investigated in \cite{Coretto_Hennig_2013_theory} and cited here.


\section{Methods from the literature}\label{sec_methods_from_literature}

In the following, assume an observed sample  $\underline{x_n}=\set{x_1, x_2, \ldots, x_n},$
where $x_i$ is the realization of a random variable $X_i \in \R^p$ with $p \geq 1$; $X_1,\ldots,X_n$ i.i.d. The goal is to cluster the sample points into $G$ distinct groups. 

{\bf Maximum Likelihood (ML) for Gaussian mixtures (gmix).~} Let $\phi(x; \mu, \Sigma)$ be the density of a multivariate Gaussian distribution with mean vector $\mu \in \R^p$ and $p\times p$ covariance matrix $\Sigma$. Assume that the observed sample  is 
i.i.d. drawn from  the finite Gaussian mixture distribution  having density 
\begin{equation}\label{eq_plain_gaussian_mixture_model}
m(x; \theta) = \sum_{j=1}^G \pi_j \phi(x;\mu_j,\Sigma_j),
\end{equation}
where $\pi_j \in [0,1]$ for all $\js$ and $\sum_{j=1}^{G} \pi_j=1$, $\theta$ is the parameter vector containing the triplets $\pi_j, \mu_j, \Sigma_j$ for all $\js$. Clustering coincides with assigning points to the mixture components based on ML parameter estimates. $\pi_j$ can be interpreted as the expected proportion of points originated from the $j$th component. Let $\theta^{\text{ml}}_n$ be the ML estimator for $\theta$, usually computed 
by the EM algorithm \citep{Dempster_Laird_etal_1977}. The ML estimator under  \eqref{eq_plain_gaussian_mixture_model} exists only under appropriate constraints on the covariances matrices. These constraints (which are also relevant for the methods introduced below) will be discussed in detail in Section \ref{sec_otrimle}. Let $\tau_{ij}^{\text{ml}}$ be the estimated  posterior probability that the observed point $x_i$ has been drawn from the $j$th mixture component, i.e., 
\begin{equation}\label{eq_bayesrule}
\tau_{ij}^{\text{ml}} = %
\frac{\pi^{\text{ml}}_{j,n} \phi(x_i; \mu^{\text{ml}}_{j,n},\Sigma^{\text{ml}}_{j,n})  }{ m(x_i; \theta^{\text{ml}}_{n}) } %
\qquad \text{for all } \quad \js.
\end{equation}
The point $x_i$ can then be assigned to the $k$th cluster if $k = {\arg\max_{\js}} \; \tau_{ij}^{\text{ml}}$.  This assignment method is common to all model-based clustering methods. gmix is implemented in R's  MCLUST  package 
\citep{Mclust_Software_Manual_2006}.
As illustrated in Section \ref{sec_introduction} and proven in \cite{Hennig_2004}, the method can be
strongly affected by outliers and deviations from the model assumptions, 
and we now turn to approaches that attempt to deal with this problem.
For lack of space, we present these in detail in the online supplement 
and only give a short overview here.

{\bf ML--type estimator for Gaussian mixtures with uniform noise (gmix.u).~} 
\cite{Banfield_Raftery_1993} added a uniform mixture component on the smallest hyperrectangle covering the data to \eqref{eq_plain_gaussian_mixture_model}, calling it ``{\it noise component}'' to accomodate ``noise''.

{\bf ML for mixtures of Student--t distributions (tmix).} \cite{McLachlan_Peel_2000b} replaced the Gaussian densities in \eqref{eq_plain_gaussian_mixture_model} with multivariate Student-t densities, because they have heavier tails and can therefore accomodate outliers in a better way. Observations can be declared ``noise'' if they lie in a low density area of the t-distribution fitting their cluster. \cite{Hennig_2004} showed that neither tmix noir gmix.u are breakdown-robust.

{\bf TCLUST} is based on maximizing a trimmed likelihood of a ``fixed partition model'' with  cluster weights $\pi_j$. With $R=\cup_{j=1}^G R_j$,  $\#\{ R \} = [n(1-\alpha)]$ the number of non-trimmed points:
\begin{equation}\label{eq_tclust_functional}
\theta^{\text{tclust}}:= \argmax_{\theta \in \Theta, \#\{ R \} = [n(1-\alpha)]}  \sum_{j=1}^G \sum_{i \in R_j} \left(\log \pi_j+\log \phi(x; \mu_j, \Sigma_j)\right).
\end{equation}
For background, see  \cite{Gallegos_2002}, \cite{Gallegos_Ritter_2005},  
\cite{GarciaEscudero_etal_2008}.  The TCLUST methodology is implemented in R's TCLUST  package by \cite{Fritz_etal_2012}. Partition methods with trimming  started with the trimmed $k$-means proposal of  \cite{Cuesta_Albertos_etal_1997}.

{\bf Further existing work.} More approaches to robust model-based clustering can be found in the literature. \cite{Neykov_Filzmoser_etal_2007} proposed and implemented a trimmed likelihood method. \cite{Qin_Priebe_2013} introduce an EM-algorithm adapted to maximum $L_q$-likelihood estimation and study its behaviour under a gross error model. References to other approaches to robust clustering
are given in \cite{GarciaEscudero_Gordaliza_2010}.


\section{Optimally tuned robust improper maximum likelihood}\label{sec_otrimle}
\subsection{Robust improper maximum likelihood}
The robust improper maximum likelihood estimator (RIMLE) is based on  the ``noise component''-idea for robustification (gmix.u).  The main idea is to use a  pseudo-model where the noise is represented  by  an improper constant density over the whole Euclidean space:
\begin{equation}\label{eq:psi}
\psi_\delta(x, \theta)=\pi_0 \delta + \sum_{j=1}^{G} \pi_j \phi(x; \mu_j, \Sigma_j),
\end{equation}
with $\pi_0,\pi_j \in [0,1]$ for $\js$, $\pi_0 + \sum\nolimits_{i=1}^G \pi_j = 1$. $\delta > 0$ is the improper constant density (icd). The parameter vector $\theta$ contains  all Gaussian parameters plus all proportion parameters including $\pi_0$, ie. $\theta = (\pi_{0},\pi_{1},\ldots,\pi_{G}, \mu_{1},\ldots,\mu_{G} ,\Sigma_{1},\ldots,\Sigma_{G}).$  Given  the sample improper pseudo-log-likelihood function
\begin{equation}\label{eq:ln}
l_n(\theta) = \frac{1}{n} \sum_{i=1}^{n} \log \psi_\delta(x_i, \theta),
\end{equation}
the RIMLE is defined as 
\begin{equation}\label{eq:rimle}
\theta_n(\delta) =   \argmax_{\theta \in \Theta} l_{n}(\theta),
\end{equation}
where $\Theta$ is an appropriate constrained parameter space discussed below. $\theta_n(\delta)$ is then used to cluster points as for model-based clustering methods. Define pseudo posterior probabilities in analogy with \eqref{eq_bayesrule}:
\begin{equation*}
\tau_{j}(x_i, \theta):= \\
\begin{cases}
  \frac{\pi_0 \delta}{\psi_\delta(x_i,\theta)}                    &   \text{if} \; j=0  \\
  \frac{\pi_j \phi(x_i,\mu_j,\Sigma_j)}{\psi_\delta(x_i,\theta)}   &   \text{if} \; \js;  \\
\end{cases}
\quad \text{for} \; i=1,2,\ldots,n,
\end{equation*}
and assign the points based on the following rule
\begin{equation}\label{eq_J}
J(x_i, \theta):=\argmax_{j\in \set{0,1,2,\ldots,G}} \tau_{j}(x_i, \theta).
\end{equation}
Fixing $\delta$, \eqref{eq:psi}   does not define a proper probability model, but \eqref{eq:rimle} yields a useful procedure for data modelled as a proportion of $(1-\pi_0)$ of a mixture of Gaussian distributions plus  a proportion of $\pi_0$ points not assigned to any meaningful cluster. Regions of high density  are rather associated with clusters than with noise, so  the noise regions should be those with the lowest density. This could be achieved by using the uniform density as in  gmix.u, but for this the presence of gross outliers the dependence of the uniform distribution on the convex hull of the data  still causes a robustness problem (\cite{Hennig_2004})
. 

The optimization  problem in \eqref{eq:rimle} requires that $\Theta$ is suitably defined, otherwise $\theta_n(\delta)$ may not exist. As discovered by \cite{Day_1969}, the Gaussian mixture likelihood can degenerate. This problem extends to \eqref{eq:psi} as well. Let $\lambda_{k,j}$ be an eigenvalue of $\Sigma_j$ for some $\ks$ and $\js$. Take a sequence $(\theta_m)_{m \in \N}$ such that $\lambda_{k,j,m}\searrow 0$ and $\mu_{j,m}=x_1$, then $l_n(\theta_m) \to +\infty$. There are various ways to avoid this issue. Let $\lmax(\theta)$ and $\lmin(\theta)$ be respectively the maximum and the minimum eigenvalues of the covariance matrices in $\theta$,  \cite{Coretto_Hennig_2013_theory} adopt the ``eigenratio constraint''
\begin{equation} \label{eq_cov_constraint}
\lmax(\theta) / \lmin(\theta)  \leq \gamma < +\infty
\end{equation}
with fixed  $\gamma\ge 1$. $\gamma=1$ constrains all component covariance matrices to be spherical and equal, as in $k$-means clustering, while $\gamma>1$ restricts the relative scatter discrepancy among  clusters. This type of constraint has been proposed by \cite{Dennis_1981} and studied by \cite{Hathaway_1985} for one dimensional Gaussian mixtures. EM-algorithms for computing the ML of  multivariate Gaussian mixtures  under  \eqref{eq_cov_constraint} have been studied by \cite{Ingrassia_2004} and \cite{Ingrassia_Rocci_2007}, although asymptotic properties  of the corresponding MLE have not been proved.  The same constraints are used for  TCLUST by \cite{GarciaEscudero_etal_2008}. There are a number of alternative constraints, see \cite{Ingrassia_Rocci_2011, Gallegos_Ritter_2009}. 

Although \eqref{eq_cov_constraint} prevents the unboundness  of the likelihood in standard mixture models and TCLUST, for RIMLE this is not enough. Points not fitted by any of the Gaussian components can still be fitted by the improper uniform component. Therefore, \cite{Coretto_Hennig_2013_theory} proposes an additional ``noise proportion constraint'',
\begin{equation} \label{eq_noise_constraint}
\frac{1}{n}  \sum_{i=1}^n  \tau_0(x_i, \theta) \leq \pi_{\text{max}},
\end{equation}
for fixed $0< \pimax < 1$. The quantity $ n^{-1} \sum_{i=1}^n  \tau_0(x_i,\theta)$ can be interpreted as the estimated proportion of noise points.  Setting  $\pimax =0.5$, just implements a familiar condition in robust statistics that at most half of the data should be classified as ``outliers/noise''.  The resulting restricted parameter space for RIMLE is then 
\begin{equation}\label{eq_Theta_n}
\Theta:= \left\{
\theta:  \; \; %
\pi_j \geq 0 \; \forall j \geq 1; \; %
\pi_0+\sum_{j=1}^G \pi_j =1; \;
\frac{1}{n}  \sum_{i=1}^n  \tau_0(x_i, \theta) \leq \pi_{\text{max}}; \; %
\frac{\lmax(\theta)}{\lmin(\theta)} \leq \gamma
\right\}. 
\end{equation}
\cite{Coretto_Hennig_2013_theory} show  that $\theta_n(\delta)$ exists for any $\delta\geq 0$ if $\#(\underline{x_n})>G+\lceil n \pimax  \rceil$ and that  $\theta_n(0)$ exists under the milder condition that $\#(\underline{x_n})>G$. For $\delta=0$, the RIMLE reduces to ML for plain Gaussian mixtures. Let  $\E_P f(x)$ be the expectation of $f(x)$ under $x \sim P$. The RIMLE functional is defined as 
\begin{equation}\label{eq_rimle_functional}
\theta^\star(\delta) = \argmax_{\theta \in \Theta_G} E_P \log \psi_\delta(x;\theta).
\end{equation}
Existence of \eqref{eq_rimle_functional}, consistency of  $\theta_n(\delta)$ on the quotient space topology identifying all loglikelihood maxima and its breakdown point are shown in \cite{Coretto_Hennig_2013_theory}. 

\subsection{Optimal improper density level}\label{sec_density_level}
Occasionally, subject matter knowledge may be available aiding the choice of $\delta$, but such situations are rather exceptional. Here we suggest a data dependent choice of $\delta$.  Note that $\delta$ is not treated as a model quantity to be estimated here, but rather as a tuning device to enable a good robust clustering. The aim of the RIMLE is to approximate the density of clustered regions of points when these regions look like those produced by a Gaussian distribution. We define the ``optimal'' $\delta$ value as minimizer  of a criterion function measuring the discrepancy of the found clusters from the Gaussian prototype. Given $\theta_n(\delta)$, define  the clusterwise squared Mahalanobis distances to clusters' centres as
\begin{equation*}
d_{i,j,n} = (x_i - \mu_{j,n})' \Sigma_{j,n}^{-1} (x_i-\mu_{j,n}),
\end{equation*}
and the clusterwise weighted empirical distribution of $d_{i,j,n}$, 
\begin{equation}\label{eq:mathbbM}
\mathbb{M}_{j}(t; \delta) = %
\frac{1}{\sum_{i=1}^n \tau_{j}(x_i, \theta_n(\delta))}  \sum_{i=1}^n  \tau_{j}(x_i, \theta_n(\delta)) \ind\{d_{i,j,n}  \; \leq \; t\},  \qquad \js.
\end{equation}
In $\mathbb{M}_{j}$, the $i$th point's distance is weighted according to the pseudo posterior probability that the $i$th observation has been generated by the $j$th mixture component. If the $j$th cluster is approximately Gaussian and $\mu_{j,n}$ and $\Sigma_{j,n}$ are good approximations of its location and scatter, we expect that squared Mahalanobis distances to $\mu_{j,n}$ of the points indeed belonging to mixture component no. $j$ (for which $\tau_j(\cdot)$ indicates the estimated probability) will approximate a $\chi_p^2$ distribution. With $\chi^2_p(a)$ being the value of the cdf of the $\chi^2_p$ distribution at $a$, define the Kolmogorov-type  distance for the $j$th cluster
\begin{equation}
\text{K}_{j}(\delta) = \max_{i=1,\ldots,n}  \abs{ \mathbb{M}_{j}(d_{i,j,n}; \delta) - \chi^2_p(d_{i,j,n})  }.
\end{equation}
The quality of the overall Gaussian approximation is then evaluated by weighting  $\text{K}_{j}(\cdot)$ with the estimated component proportion $\pi_{j,n}$:
\begin{equation}\label{eq_orimle_obj}
\text{D}(\delta) = \frac{1}{\sum_{j=1}^G \pi_{j,n}} \sum_{j=1}^G \pi_{j,n} \text{K}_{j}(\delta). 
\end{equation}
For a given constant $\beta \geq 0$, define  the optimal icd level as 
\begin{align}\label{eq_orimle}
\delta_n = \argmin_{\delta \in [0, \delta_{\max}]} \; \text{D}(\delta) + \beta \pi_{0,n}.
\end{align}
The corresponding optimally tuned RIMLE (OTRIMLE) will be denoted $\theta_n(\delta_n)$. Existence and uniqueness of $\delta_n$ are not trivial, see Section \ref{sec_computing}. Section \ref{sec_delta} is about how $D(\delta)$ behaves as a function of $\delta$.

The straightforward choice for $\beta$, formalizing the ``Gaussian cluster'' concept, is $\beta=0$. However, often in practice it is not so important that the clusters are of as precise Gaussian shape as possible. $\beta>0$ (but normally smaller than 1) formalizes that less Gaussian shapes of clusters are tolerated if this brings the estimated noise proportion down. As can be seen in Section \ref{sec_simulation_study}, choosing $\beta=\frac{1}{3}$ leads to improvements if the true clusters are $t$-distributed. Section \ref{sec_beta} gives more details on the effect of choosing $\beta>0$.


\subsection{Computation}\label{sec_computing}
For a fixed $\delta$ the RIMLE can be appropriately computed using an Expectation--Maximization (EM)-algorithm. See \cite{Coretto_Hennig_2013_theory} for details, and the online supplement 
for details and background of the following. The outcome of the EM-algorithm depends on the initialization. 
We used an initialization method inspired by the MCLUST software. 
In order to avoid spurious clusters, we consider as valid initial partitions  only those containing at least  {\tt min.pr}$\times n$ observations in each cluster ({\tt min.pr}=$0.005$, say). As a first attempt to find such a valid partition, nearest neighbor based clutter/noise detection proposed by  \cite{Byers_Raftery_1998} is applied to identify an initial noise guess. Agglomerative hierarchical clustering based on ML  criteria for Gaussian mixture models proposed by \cite{Banfield_Raftery_1993} is then  used for finding initial Gaussian clusters among the non-noise. See the online supplement for the case that the found partition is not ``valid''.

The OTRIMLE can be found by computing RIMLEs on a grid of $\delta$ values ranging from zero to some large enough $\delta_{\max}$. In practice, we solve the program \eqref{eq_orimle} by the ``golden section search'' of  \cite{Kiefer_1953} over the candidate set $\delta\in [0,\delta_{\max}]$.  In most numerical experiments we found that no more than 30 RIMLE evaluations are required. $\delta_{\max}$ can be chosen as highest density value occurring within an initialized cluster, discarding $\delta$-values for which the RIMLE-solution ends up at the border of the parameter space.


\section{Definition of ``true'' clusters and misclassification}\label{sec_true_clusters}

In most simulation studies in cluster analysis, in which data are generated from mixture (or fixed partition) models, it is assumed that the ``true'' clusters are identified with the mixture components, and methods can then be evaluated by misclassification rates. But this can be problematic. Consider the comparison of ML-estimators for Gaussian mixtures and for mixtures of t-distributions. In most applications, both approaches would be considered as potentially appropriate for doing the same thing, namely finding clusters that are unimodal and elliptical. In applications in which the clustering is of main interest (as opposed to parameter estimation), researchers would not mind much whether the density around their cluster cores rather looks like a Gaussian or a t-distribution. But implications for which points are considered as ``true outliers'' vs. ``truly belonging to a cluster'' would be different, because some points generated by a t-distribution with low degrees of freedom are indeed outlying with respect to the core of the t-distribution from which they are generated. 

More generally, the identification of clusters and mixture components cannot be taken for granted. \cite{Hennig_2010} illustrates that the interpretation of Gaussian mixture components as clusters depends on whether the components are separated enough. And in robust clustering, one would often interpret a group of a few points with low density as ``noise'' even if they were generated by a Gaussian distribution.

\begin{figure}[!t]
\centering
\includegraphics[width=\textwidth]{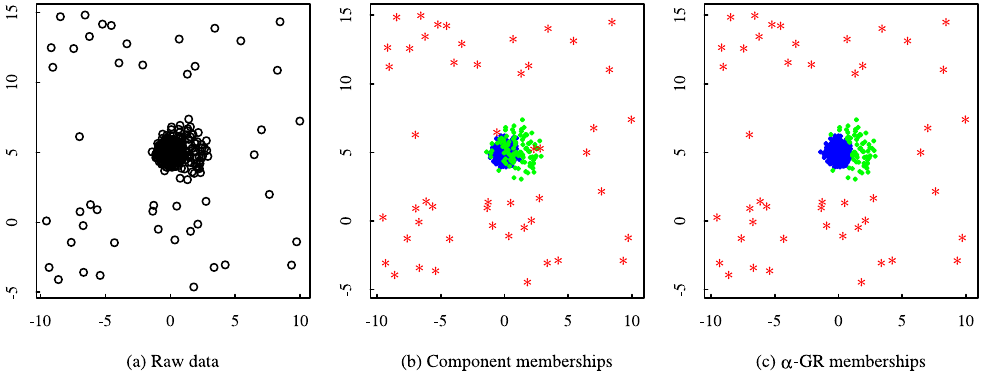}
\caption{An artificial dataset consisting of 950 points drawn from two 2-dimensional Gaussian distributions and 50 points from a uniform distribution on the square $[-10,10]\times[-5,15]$. (a) unlabeled points, (b) colored according to the mixture components, (c) colors represent the two GR$_\alpha$ (with $\alpha=10^{-4}$); red stars belong to NR$_\alpha$. }%
\label{fig_true_cluster_example}
\end{figure}

We now define a ``reference truth'' for the mixture models that are used in our simulation study, from which misclassification rates then can be computed. For motivation consider Figure \ref{fig_true_cluster_example}, which shows an artificial dataset drawn from a mixture of two Gaussians and a uniform distribution. Figure \ref{fig_true_cluster_example}(a) shows the unlabeled dataset on which cluster analysis operates. Figure \ref{fig_true_cluster_example}(b) shows the points labelled by the mixture components that generated them. Observe  that there are three red stars (generated from the uniform noise) in the middle of the region where the two Gaussians have most of their mass. Furthermore, there are green  points from the left Gaussian component with lower density that fall into the dense blue region. No method can be expected to reconstruct all the cluster memberships in such overlapping regions.
Figure \ref{fig_true_cluster_example}(c) shows what we define as the ``reference truth'', defined next.

The idea is that we choose probability measures $P_1,P_2,\ldots,P_G$ to correspond to the $G$ ``true clusters'' (implying that they ``cluster'', i.e., generate clearly distinguishable, although not necessarily non-overlapping, data patterns). For each of these, we define a region of points that can be considered as non-outliers based on a mean and covariance matrix functional, which are Fisher-consistent at the Gaussian distribution, but robust and existing for other distributions, too. This defines a region of non-outliers of Gaussian shape. We consider all points ``noise'' that are outliers according to this definition for all $P_1,P_2,\ldots,P_G$. Quadratic discriminant analysis assigns points to clusters that are non-outliers with respect to more than one of the $P_j$. This means that points are assigned to clusters by optimal classification boundaries under the Gaussian assumption, even if the components are in fact not Gaussian. This formalizes using ``Gaussian cluster prototypes'' without assuming that clusters really have to be Gaussian. 
 
To this end, let $m_j$ and $S_j$ be the Minimum Covariance Determinant (MCD) center and scatter functional at  $P_j$ \citep{Rousseeuw_1985}. \cite{Cator_Lopuhaa_2012} proved existence of the MCD functional for a wide class of probability measures.   The  $S_j$ can be corrected for achieving consistency at $P_j$ equal to the Gaussian distribution,  \citep{Croux_Haesbroeck_1999, Pison_VanAelst_2002}, so that  when $P_j$ is Gaussian,  $m_j$ and $S_j$ are the corresponding  mean vector and covariance matrix. Let $\pi_j$ be the expected proportion of  points generated from $P_j$.  We allow $\sum_{j=1}^G p_j \leq 1$, so that  points could be generated by (noise-)distributions other than $P_1,P_2,\ldots,P_G$. Define the quadratic discriminant score for assigning the point $y \in \R^p$ to the $j$th cluster by maximizing  
\begin{equation*}
\text{qs}(y; \pi_j, m_j, S_j):=  \log (\pi_j)  -  \frac{1}{2} \log (\det(S_j)) -  \frac{1}{2}(y-m_j)' S_j^{-1} (y-m_j),  \qquad \text{for} \; j=1,2,\ldots,G.
\end{equation*}
If clusters are indeed Gaussian, this is equivalent to \eqref{eq_J}.
Consider 
\begin{equation*}
\mathcal{E}_{\alpha}(m_j, S_j):= \{y: \;\; (y-m_j)' S_j^{-1} (y-m_j) \leq  \chi^2_{p}(1-\alpha) \},
\end{equation*}
where $\chi^2_p(1-\alpha)$ is the $1-\alpha$ quantile of the $\chi^2_p$ distribution. For a fixed $\alpha$, the ellipsoid $\mathcal{E}_{\alpha}(m_j, S_j)$ defines the subset of $\R^p$ that hosts the $j$th cluster. The size of this ellipsoid is defined in terms of $\chi^2_p(1-\alpha)$, because for Gaussian $P_j$, $P_j (\R^p \setminus  \mathcal{E}_\alpha(m_j, S_j))=\alpha$. For a fixed level $\alpha$, the $\alpha$-Gaussian Region is defined as the union of these ellipsoids:  
\begin{equation*}\label{eq_GRalpha}
\text{GR}_{\alpha} := \bigcup_{j=1}^G \mathcal{E}_{\alpha}(m_j, S_j), 
\end{equation*}
and the noise region is given by $\text{NR}_{\alpha} := \R^p \setminus \text{GR}_{\alpha}$.
\begin{definition}[$\alpha-$Gaussian cluster memberships]\label{definition_alpgaGR}
Given $\alpha \in [0,1)$, a data generating process (DGP) with cluster parameters $\theta_C:=\{(\pi_j, m_j, S_j),\; \js\}$, and a dataset $\underline{x_n}:=\{x_1,x_2,\ldots,x_n\}$; the  $\alpha-$Gaussian cluster memberships are given by
\begin{equation}\label{eq_agr}
\text{AGR}_\alpha(x_i;\theta_C) := \ind\{x_i \in \text{GR}_{\alpha}\} \times \argmax_{\js} \:\: j\:\ind\{j={\argmax}_{g=1,\ldots,G}  \;\; \text{qs}(y; \pi_g, m_g, S_g) \}.
\end{equation} 
\end{definition}
$\text{AGR}_\alpha(x_i, \theta_C)=0$ means that $x_i \in \text{NR}_{\alpha}$.  

This definition is inspired by the definition of outliers with respect to a reference model as in \cite{Davies_Gather_1993},  \cite{Becker_Gather_1999}. A difference is that here the parameter $\alpha$ does not directly control the probability of the noise region. Once $\alpha$ and the triples $(\pi_j, m_j, S_j)$ are fixed for all $\js$, the size of the noise region will depend on the degree of overlap and Gaussianity of the ellipsoids in $\text{GR}_{\alpha}$. $\alpha$ needs to be small because the idea of an outlier implies that under the Gaussian distributions outliers are very rare. We choose $\alpha=10^{-4}$, which implies that the probability that there is at least one outlier in $n=500$ i.i.d. Gaussian observations is $0.0488$.  

The different robust clustering methods have different implicit ways of classifying points as ``noise'' (noise component, trimming, outlier identification in t-distributions). In order to make them comparable, we use \eqref{eq_agr} to unify the point assignment of the methods by computing $\text{AGR}_\alpha(\cdot)$ based on the parameters estimated by the methods, from which we assume that estimators of the triples $(\pi_j, m_j, S_j)$ (cluster proportion, center, and covariance matrix) can be computed (see Section \ref{sec_methods_under_comparison}). Let the estimated cluster parameters be $\hat \theta_{C,n}$. A misclassification rate can then be computed by applying an optimal permutation $\sigma\{\cdot\}$ of cluster labels:
\begin{equation}\label{eq_mcr}
\text{mcr}(\theta_{C},\hat \theta_{C,n})  = \argmin_{\sigma}   \frac{1}{n} \sum_{i=1}^n \ind\{ \text{AGR}_\alpha(x_i; \theta_{C}) = \sigma\{\text{AGR}_\alpha(x_i; \hat \theta_{C,n})\} \}.
\end{equation}
See the online supplement 
for computation of the MCD for non-normal distributions.


\section{A comparative simulation study}\label{sec_simulation_study}
Here we present a comprehensive simulation study comparing the OTRIMLE with the methods introduced in Section \ref{sec_methods_from_literature}.

\subsection{Data generating processes}\label{sec_dgp}
The methods are compared on a total of 24 DGPs with 1000 Monte Carlo replicates each. Half of the DGPs produce 2-dimensional datasets. The remaining twelve DGPs are 20-dimensional versions that are constructed adding independent 18-dimensional uncorrelated zero-means unit-variance Gaussian and/or Student-t marginals. 
Therefore clusters are always only defined on the first two marginals. Note that the aim of the simulation study is not variable selection; we designed the DGPs so that clustering information is only in the first two dimensions in order to be able to visualize and control the clustering patterns, but we compare clustering methods that use all variables (trying variable selection methods is beyond the scope of this paper). We do not think that variable selection or dimension reduction is mandatory in clustering, because the meaning of the clusters is determined by the involved variables, see the discussion rejoinder in \cite{Hennig_Liao_2013}.
%
We choose $n=1000$ for 2-dimensional designs and $n=2000$ for the 20-dimensional versions. DGPs have been designed in order to test a variety of ``noise patterns'', numbers of clusters $G$, and patterns of separation/overlap between different clusters.

We consider two main classes of DGP, namely DGPs with a uniform noise component on the first two marginals,  and DGPs that do not have a noise component. The first group includes the following setups, all of which have clusters generated from Gaussian distributions, and for $p=20$ the 18 uninformative variables are Gaussian:  (i) for ``WideNoise'' DGPs, the uniform noise component produces points that are widespread but overlap with the  clustered regions entirely; (ii) for ``SideNoise'' DGPs the uniform noise component spreads points on a wide region that overlaps slightly with some of the clusters; (iii)  in ``SunSpot'' DGPs there is a uniform component that produces few extremely outlying points. On the other hand we consider DGPs that do not include a noise component (i.e. $\pi_0=0$). This second group can be divided into three further subgroups: (i)/(ii) in ``GaussT'' and ``TGauss'' DGPs, multivariate Student-t distributions  with three degrees of freedom are used. In ``GaussT'' these are used as uninformative distributions for $p=20$, whereas the first two clustered dimensions use Gaussians; in ``TGauss'' the clusters are generated by noncentral multivariate $t_3$-distributions and for $p=20$ the 18 uninformative variables are Gaussian; (iii) in ``Noiseless'' DGPs,  all points are drawn from Gaussian distributions. 

For each of the six setups, there are variants with a lower and a higher number of clusters $G$, $p=2$ (denoted by ``l'') and $p=20$ (denoted by ``h''), adding up to 24 DGPs. The nomenclature used in the following puts these at the end of the setup name, i.e., ``TGauss.5h'' refers to the ``TGauss''-setup with higher $G=5$ and higher $P=20$.
For ``WideNoise'', ``SideNoise'', and ``GaussT'', the lower $G$ was 2 and the higher $G$ was 3. For ``SunSpot'', ``TGauss'' and ``Noiseless'', the lower $G$ was 3 and the higher $G$ was 5. The overlap between clusters as well as the combinations of cluster shapes varied between DGPs.
Full details of the definition of the DGPs are given in the online supplement
together with exemplary scatterplots of the first two dimensions of a dataset from every setup. 
 
\subsection{Implementation of methods}\label{sec_methods_under_comparison}
Table \ref{tab_methods} summarizes settings for the compared methods.  TCLUST and RIMLE/OTRIMLE are based on eigenratio constraints, but  this is not the case for the MCLUST software \citep{Mclust_Software_Manual_2006} and the available implementation of mixtures of t-distributions \citep{McLachlan_Peel_2000b}. In order to have full comparability of the solutions, the eigenratio constraints (see Section \ref{sec_computing}) have been implemented by us for OTRIMLE/RIMLE, gmix, tmix and gmix.u; the latter is computed by use of the same routine that is used for RIMLE /OTRIMLE. For TCLUST, the constraints are in the original R-package. 

For all methods the eigenratio constraint has been set equal to 20 for each of the 24 DGPs. The latter choice is motivated by the fact that 20 is larger than the maximum true eigenratio across the designs involved in the comparison, and it is in general a value that often enables rather smooth optimization (obviously in reality the true eigenvalue ratios are not known, but for variables with comparable scales and value ranges, 20 gives the covariance matrices enough flexibility for most applications). OTRIMLE has been tested with and without the penalty term $\beta=\frac{1}{3}$ in \eqref{eq_orimle}, denoted by ot.rimle ($\beta=0$) and ot.rimle.p respectively. Other values for $\beta$, ranging from 0.1 to 0.5,  have been tried, and results did not change much. A difficulty with TCLUST is that an automatic data driven choice of the trimming level is currently not available. In tclust.f we  set the trimming level to 10\%. This  choice is motivated by the fact the DGPs  produced an average  proportion of points belonging to the NR$_\alpha$ set  in the range [0\%, 23\%]  \citep[see Table 1 in][]{Coretto_Hennig_2014_supplement}. Furthermore, since the trimming level plays a role similar to $\delta$ in RIMLE/OTRIMLE, there are two versions of TCLUST for which the same idea for automatic  decisions  the trimming level is used as proposed here for OTRIMLE, see Section \ref{sec_otrimle}. In ot.tclust and ot.tclust.p the trimming level has been selected using  \eqref{eq_orimle} with  the trimming level playing  the role of  $\delta$, and the   weights $\tau_j(\cdot)$ in $\mathbb{M}_j(\cdot)$ replaced by the 0-1 crisp weights of TCLUST, again using $\beta=0$ or $\beta=\frac{1}{3}$. For t-mixtures, in the original proposal by  \cite{McLachlan_Peel_2000b}, the degrees of freedom are assumed to be equal across the mixture components and are estimated from the data and covariance matrix constraints are not considered.   In tmix we fix the degrees of freedom to 3 and we incorporate eigenratio constraints, as for the other methods. This is motivated as follows: (i) for some of the DGPs (particularly the SunSpot designs) constraints were needed in order to avoid  spurious solutions; (ii) for some of the sampling designs not based on  Student-t distributions, estimation of the degrees of freedom of the t-distribution produced an extremely large   variability  in the misclassification rates; (iii) since a number of designs are based on Student-t with 3 degrees of freedom, the decision to fix this parameter to 3 gives the t-mixture a slight advantage, which seems fair given that the majority of setups rather seem to favour noise component/trimming. \\ 

For some of the DGPs the experience suggested that solutions may depend strongly on the initialization. In order to reduce the bias introduced by different initializations, all methods  are initialized from the same partition, see Section \ref{sec_computing} (an additional set of R functions has been provided by TCLUST's authors to allow for this). 


\begin{table}[t]
  \centering
  \caption{Summary of the main settings for the methods under comparison.}
  \label{tab_methods}
  \begin{tabular}{lp{0.82\textwidth}}
    \toprule
    {\bf Method}       & {\bf Setup} \\ 
    \midrule
    {\bf gmix.u}       & RIMLE with  $\delta=1/V_n$ where $V_n$ is volume of the smallest hyperrectangle that contains the data. \\ 
    {\bf ot.rimle}     & OTRIMLE without penalty ($\beta=0$). \\ 
    {\bf ot.rimle.p}   & OTRIMLE with penalty term $\beta=1/3$.\\ 
    {\bf tclust}       & TCLUST with fixed trimming level set at 0.1. \\ 
    {\bf ot.tclust}    & TCLUST with trimming level selected by the OTRIMLE criterion without penalty ($\beta=0$).\\ 
    {\bf ot.tclust.p}  & TCLUST with trimming level selected by the OTRIMLE criterion with penalty term $\beta=1/3$. \\
    {\bf tmix}         & ML for the Student-t mixture model \eqref{eq_plain_gaussian_mixture_model}  with $v=3$ for all components plus eigenratio constraints. \\
    {\bf gmix}         & ML for the Gaussian mixture model  plus eigenratio constraints. \\
    \bottomrule
  \end{tabular}
\end{table}


\subsection{Results}\label{sec_results}
\begin{figure}[t]
\centering
\includegraphics[width=\textwidth]{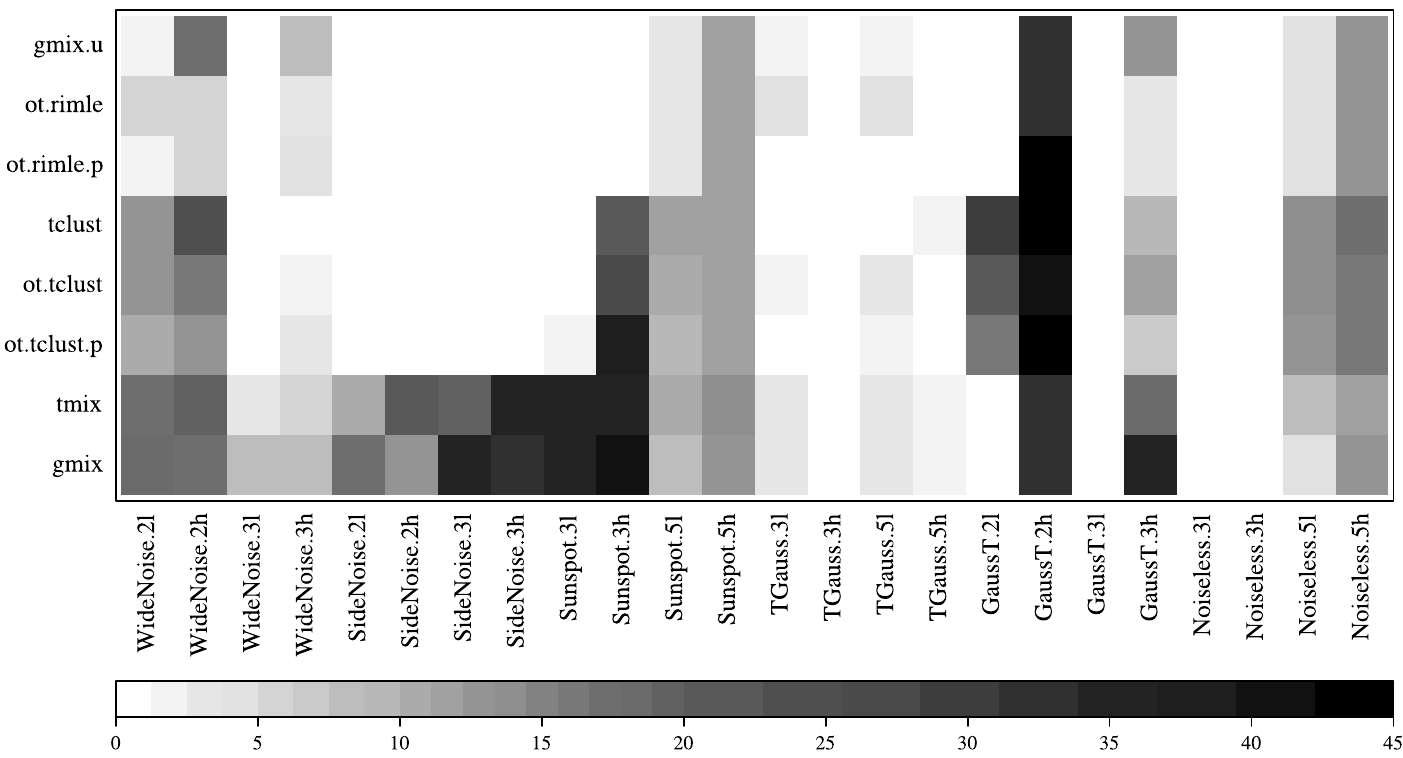}
\caption{Level plot representing the  sample mean of the Monte Carlo distribution of misclassification rates (percentage scale) for each DGP-method pair. Each square of the plot represents the average misclassification according to the bottom grey color scale. }%
\label{fig_mcr_levelplot}
\end{figure}
The methods are compared using misclassification rates as defined in  \eqref{eq_mcr}. These are more relevant in clustering tasks than parameter estimates. The results are graphically  summarized in  Figure \ref{fig_mcr_levelplot}, while average misclassification rates with standard errors are given in Table \ref{tab_global}.  Each square in the plot is a color-coded representation of the misclassification rate averaged over the 1000 Monte Carlo replicates for a given method-DGP pair.  
Further details about  average  misclassification rates are given in the online supplent. 
It also contains boxplots of the misclassification rates for all method-DGP pairs. Figure \ref{fig_mcr_levelplot} shows clear evidence that using robust methods is important. The gmix method only performs well for Noiseless and some DGPs with Gaussians and t-distributions, but most other methods work well (although slightly worse at times) for these DGPs, too. tmix works well for most DGPs involving t-distributions, but for the other DGPs with noise/outliers, it is often seriously worse than gmix.u, OTRIMLE and TCLUST.

gmix.u performs relatively well, although for a number of DGPs it suffers strongly from high dimensionality. For WideNose in 2 dimensions, gmix.u will equal in many cases a proper ML-estimator, so the method should be advantageous here, and gmix.u is indeed best for these DGPs. However, for the 20-dimensional WideNoise, taking a uniform distribution over the smallest hyperrectangle containing the data can no longer be the ML estimate for the noise-generating mixture component, and its performance deteriorates strongly. 

Regarding the TCLUST methods,  even though the automatic trimming level of ot.tclust and ot.tclust.p did not always improve the results, it demonstrated to provide a reasonable choice for the trimming level. In fact,  for all  situations where the true average noise proportion is about equal to the trimming level of tclust, performances of tclust, ot.tclust and ot.tclust.p are very similar, meaning that the OTRIMLE criterion is a good  starting point for fixing the trimming rate.
Compared to the RIMLE-type methods, the performance of TCLUST suffers in situations where there is a considerable degree of overlap between clusters. For DGPs with overlap (such as WideNoise.2, SunSpot.5, GaussT.2 and Noiseless.5) the misclassification rate of TCLUST is completely dominated by misclassifications  between clusters (to see this consider Table \ref{tab_global} and Table 4 in the online supplement). 
The reason is that the TCLUST parameters are based on a classification-type likelihood, which relies on the separation between clusters. The TCLUST also seems to not tolerate the large number of Student-t marginals of GaussT.2h and GaussT.3h. TCLUST performs well for a number of DGPs and is clearly best in WideNoise.2h.

The OTRIMLE methods show a very good overall performance. They produce high misclassification rates only for some 20-dimensional DGPs for which all methods are in trouble (performances for GaussT.2h are generally bad with even tmix, the best method there, producing an average misclassification rate of more than 30\%), and they are best for a number of DGPs, particularly 20-dimensional WideNoise and some TGauss-DGPs. The comparison between ot.rimle and ot.rimle.p is mixed (as between ot.tclust and ot.tclust.p), with $\beta=\frac{1}{3}$ improving matters clearly for TGauss.2l and TGauss.3l (the shape of the t-distribution encourages ot.rimle to assign too many points to the noise; see Tables 2 and 3 in the online supplement,
but being significantly worse for WideNoise.3h. 

Comparing OTRIMLE with gmix.u, there are a number of DGPs for which gmix.u has a slightly lower misclassification rate than one or both of ot.rimle and ot.rimle.p. In all of these DGPs, all of gmix.u, ot.rimle and ot.rimle.p basically produce the same clustering structure, with disagreement only about the classification of some borderline points. Differences are more substantial in the setups in which gmix.u is worse. For WideNoise.2h, gmix.u in most cases does not detect any noise, so that one of the clusters consists mainly of noise. For WideNoise.3h, sometimes all or much noise is merged into a cluster, with some impact on the clustering structure. For GaussT.3h, substantial amounts of outliers are integrated into the clusters.

\begin{table}[!t]
  \begin{center}
    \caption{Monte Carlo average misclassification rates (\%)  with their standard errors in brackets. Misclassification rates are computed as in \eqref{eq_agr}. Notice that both averages (and  standard errors) are reported in percentage scale.}
    \label{tab_global}
    \resizebox{\textwidth}{!}{
      \begin{tabular}{lrrrrrrrr}
        \toprule
        {DGP} & \multicolumn{8}{c}{Method}\\
        \cmidrule{2-9}
        {}    & gmix.u & ot.rimle &  ot.rimle.p &  tclust &  ot.tclust & ot.tclust.p & tmix & gmix \\
        \midrule
        WideNoise.2l & 1.33(0.02) & 5.00(0.29) & 1.35(0.02) & 12.70(0.33) & 13.11(0.32) & 10.20(0.32) & 17.48(0.10) & 18.40(0.04) \\ 
        WideNoise.2h & 17.36(0.04) & 5.28(0.05) & 5.40(0.06) & 22.94(0.32) & 16.20(0.33) & 12.85(0.25) & 18.90(0.03) & 17.43(0.04) \\ 
        WideNoise.3l & 0.34(0.01) & 0.42(0.01) & 0.40(0.01) & 0.44(0.01) & 0.38(0.01) & 0.39(0.01) & 3.10(0.02) & 7.79(0.13) \\ 
        WideNoise.3h & 7.59(0.18) & 3.18(0.09) & 4.27(0.09) & 0.72(0.01) & 1.31(0.02) & 3.04(0.06) & 5.91(0.04) & 8.56(0.15) \\ 
        SideNoise.2l & 0.01(0.00) & 0.03(0.01) & 0.04(0.02) & 0.18(0.01) & 0.02(0.00) & 0.01(0.00) & 10.76(0.11) & 16.68(0.16) \\ 
        SideNoise.2h & 0.03(0.01) & 0.09(0.02) & 0.15(0.03) & 0.20(0.01) & 0.20(0.01) & 0.66(0.02) & 20.03(0.03) & 12.61(0.15) \\ 
        SideNoise.3l & 0.06(0.00) & 0.08(0.01) & 0.11(0.03) & 0.19(0.01) & 0.07(0.00) & 0.06(0.00) & 18.89(0.47) & 36.60(0.40) \\ 
        SideNoise.3h & 0.13(0.00) & 0.19(0.03) & 0.24(0.04) & 0.25(0.01) & 0.21(0.01) & 0.53(0.02) & 34.23(0.16) & 31.59(0.31) \\ 
        Sunspot.3l & 0.11(0.00) & 0.12(0.00) & 0.11(0.00) & 0.21(0.00) & 0.13(0.01) & 2.17(0.36) & 34.65(0.11) & 35.38(0.19) \\ 
        Sunspot.3h & 0.24(0.00) & 0.24(0.00) & 0.24(0.00) & 21.38(0.95) & 27.19(1.00) & 36.83(1.01) & 34.75(0.06) & 40.00(0.37) \\ 
        Sunspot.5l & 3.32(0.12) & 3.39(0.12) & 3.37(0.12) & 11.25(0.13) & 10.01(0.14) & 9.53(0.14) & 10.01(0.10) & 7.92(0.16) \\ 
        Sunspot.5h & 11.72(0.10) & 11.63(0.10) & 11.65(0.10) & 11.88(0.10) & 11.63(0.10) & 11.61(0.10) & 13.84(0.12) & 13.30(0.13) \\ 
        TGauss.3l & 1.84(0.02) & 4.56(0.09) & 0.90(0.01) & 0.81(0.01) & 1.63(0.03) & 0.95(0.01) & 3.28(0.02) & 3.33(0.06) \\ 
        TGauss.3h & 0.49(0.01) & 0.53(0.01) & 0.51(0.01) & 0.77(0.01) & 0.51(0.01) & 0.49(0.01) & 0.83(0.01) & 0.85(0.02) \\ 
        TGauss.5l & 1.77(0.02) & 4.65(0.07) & 1.20(0.02) & 1.21(0.01) & 2.55(0.04) & 1.32(0.02) & 3.37(0.02) & 3.59(0.05) \\ 
        TGauss.5h & 0.92(0.02) & 0.97(0.01) & 0.94(0.01) & 1.39(0.01) & 0.96(0.01) & 0.89(0.01) & 1.43(0.02) & 1.65(0.04) \\ 
        GaussT.2l & 0.57(0.01) & 0.57(0.01) & 0.56(0.01) & 28.77(0.33) & 21.50(0.48) & 16.52(0.48) & 0.83(0.02) & 0.57(0.01) \\ 
        GaussT.2h & 32.45(0.05) & 33.60(0.14) & 44.91(0.10) & 42.42(0.13) & 41.72(0.15) & 43.13(0.13) & 31.32(0.16) & 33.17(0.03) \\ 
        GaussT.3l & 0.11(0.00) & 0.11(0.00) & 0.11(0.00) & 0.27(0.01) & 0.13(0.00) & 0.12(0.00) & 0.17(0.01) & 0.11(0.00) \\ 
        GaussT.3h & 12.32(0.12) & 3.69(0.03) & 3.44(0.03) & 9.87(0.03) & 12.22(0.04) & 6.85(0.08) & 18.34(0.03) & 36.23(0.49) \\ 
        Noiseless.3l & 0.12(0.00) & 0.12(0.00) & 0.12(0.00) & 0.28(0.01) & 0.15(0.01) & 0.13(0.00) & 0.18(0.01) & 0.12(0.00) \\ 
        Noiseless.3h & 0.25(0.00) & 0.25(0.00) & 0.25(0.00) & 0.38(0.00) & 0.28(0.00) & 0.28(0.00) & 0.29(0.00) & 0.25(0.00) \\ 
        Noiseless.5l & 4.53(0.08) & 4.47(0.08) & 4.43(0.08) & 14.43(0.13) & 13.33(0.13) & 13.14(0.14) & 8.23(0.14) & 4.58(0.09) \\ 
        Noiseless.5h & 12.39(0.10) & 12.63(0.10) & 12.62(0.10) & 16.69(0.12) & 16.13(0.12) & 16.10(0.12) & 11.92(0.08) & 12.37(0.10) \\ 
        \bottomrule
      \end{tabular}
    }
  \end{center}
\end{table}

\subsection{Behavior of $D(\delta)$}\label{sec_delta}
\begin{figure}[!t]
  \centering
  \includegraphics[width=0.49\textwidth]{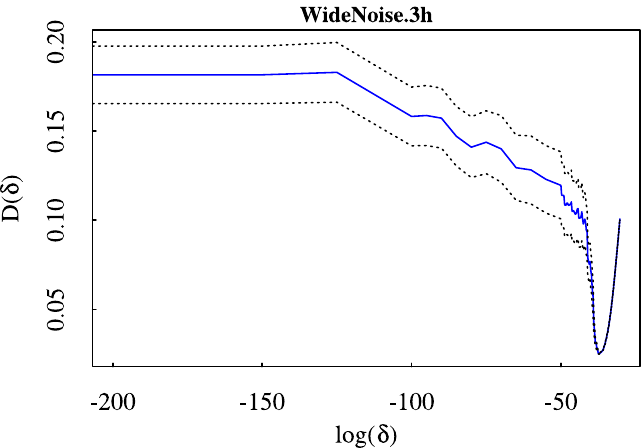}
  \hfill 
  \includegraphics[width=0.49\textwidth]{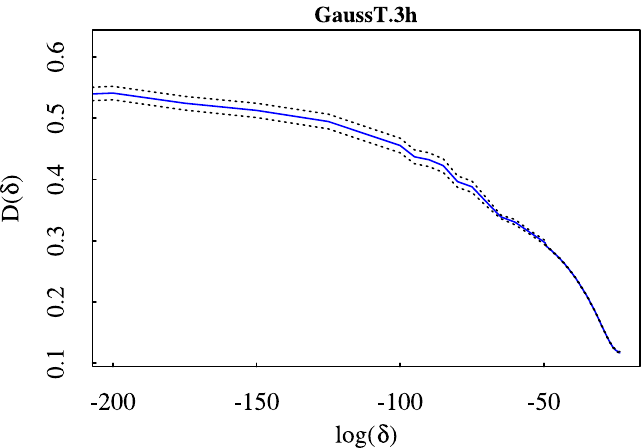}
  \caption{Monte Carlo average for OTRIMLE criterion D$(\delta)$ (blue solid line) $\pm$ standard errors (dotted lines) computed over a grid of values for $\delta \in \{0\} \cup [2.22{\times}10^{-308}, 1]$. The two plots refer to DGPs ``WideNoise.3h'' and ``GaussT.3h'' in Section \ref{sec_simulation_study}.}
  \label{fig_logicd_criterion}
\end{figure}

Here we investigate the behavior of $\text{D}(\delta)$ as a function of $\delta$ via Monte Carlo experiments under various DGPs from Section \ref{sec_simulation_study}. For each of these DGPs we produced 100 independent samples. We computed $\text{D}(\delta)$ for a grid of $\delta$ values taken from an interval  $[2.22{\times}10^{-308}, 1]$,  adding  $\delta{=}0$. In Figure \ref{fig_logicd_criterion} we report Monte Carlo averages$\pm$standard errors for $\text{D}(\delta)$ for two selected DGPs: WideNoise.3h and GaussT.3h. These are defined in Section~\ref{sec_dgp}, both with $p{=}20$.
The main difference is that noise is produced by a two dimensional uniform distribution in WideNoise.3h, whereas in GaussT.3h dimensions 3-20 are from centered $t_3$ distributions, generating some outliers. Figure \ref{fig_logicd_criterion} reports the behavior of $\text{D}(\delta)$ for $\log(\delta)>-200$. For smaller values of $\delta$ (including $\delta{=}0$) the behaviour of the curves was basically constant. 
In both graphs there is a clear minimum, although for GaussT.3h this minimum lies on the border. For WideNoise.3h the OTRIMLE criterion has a nice convex behaviour around its minimum. 
In GaussT.3h, in dimensions 3-20 the distributional shape of the clusters deviated from Gaussianity by heavier tails, although the core of the distribution looks similar to a Gaussian. D$(\cdot)$ then enforces a Gaussian shape by assigning many points to the noise component. The result is that $\pi_{0,n}$ becomes large, and the optimal $\delta$ happens at point where larger values of $\delta$ do not produce parameter estimates within the constrained set anymore (unless the constraint is enforced). The latter is a reason for the use of the penalty term in \eqref{eq_orimle} (see also Section \ref{sec_beta}). For the remaining 22 DGPs of Section \ref{sec_dgp} we found similar patterns (mostly similar to Widenoise.3h here). Another observation in Figure \ref{fig_logicd_criterion} is that around the minimum $D(\delta)$ seems to be quite stable for different datasets from the same design.

\subsection{Effect of $\beta$} \label{sec_beta}
\begin{figure}[!t]
  \centering
  \includegraphics[width=0.49\textwidth]{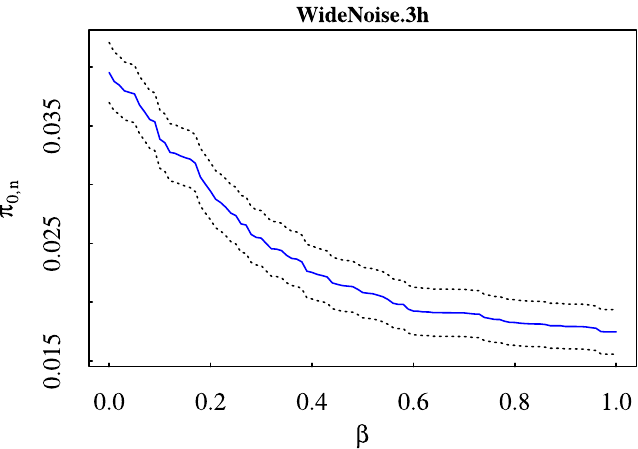}
  \hfill
  \includegraphics[width=0.49\textwidth]{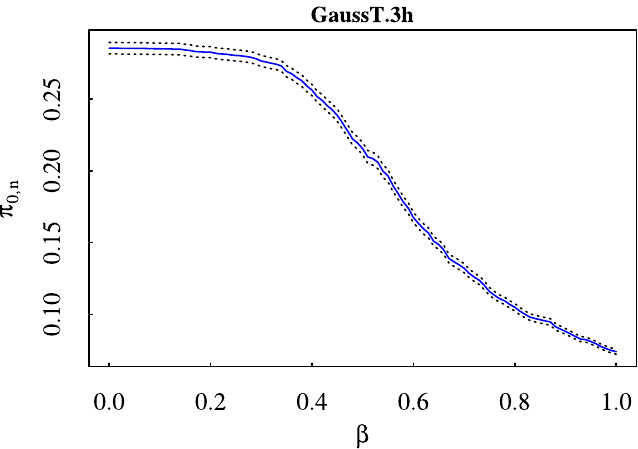}
  \caption{Monte Carlo average for the estimated noise proportion $\pi_{0,n}$ (blue solid line) $\pm$ standard errors (dotted lines) computed over a grid of values for $\beta \in [0,1]$. The two plots refer to DGPs called ``WideNoise.3h'' and ``GaussT.3h'' in Section \ref{sec_simulation_study}.}%
  \label{fig_beta_pinot}
\end{figure}

For all 24 DGPs considered in Section \ref{sec_dgp} we also investigated the behaviour of the noise proportion $\pi_{0,n}$ as a function of $\beta$, see \eqref{eq_orimle}. For each design we produced 100 independent samples, and for each of these we computed the OTRIMLE solution $\theta_n(\delta_n)$ for various values of $\beta \in [0,1]$. Figure \ref{fig_beta_pinot} reports the Monte Carlo average of the estimated noise proportion $\pi_{0,n} \pm$ standard error. For both WideNoise.3h and GaussT.3h, an increase in $\beta$ reduces smoothly the estimated noise proportion. There is some difference, however, in scales. The impact of $\beta$ is much stronger for GaussT.3h, and the same happens for all those sampling designs in which within-cluster distributions deviate from Gaussianity. Results for other DGPs are quite similar.

The discovery of the overall clustering structure was never affected by changing $\beta$ between 0 and 0.5 for the DGPs from Section \ref{sec_simulation_study}.  In the majority of cases there was no change of the clustering at all. The only difference was that larger $\beta$ sometimes produced a lower percentage of data classified as noise. In the case that $t_3$-distributions were involved, this was good, because with $\beta=0$ only fairly small central cores (60--70\%) of the points generated by t-distributions were assigned to the clusters,
whereas for larger $\beta$ only points were classified as noise that were really quite ``outlying''. On the other hand, with a larger $\beta$, in data from DGPs with Gaussian clusters and some noise that was not clearly separated from the clusters, more ``true'' noise points were assigned to the clusters.  There is no objective rule for what percentage of points from a t-distribution should be considered as ``truly'' outlying, 
and therefore it needs to be decided by the user how ``tolerant'' OTRIMLE is desired to be toward heavier distributional tails than the Gaussian ones. 

\begin{figure}[!t]
  \centering
  \includegraphics[width=0.49\textwidth]{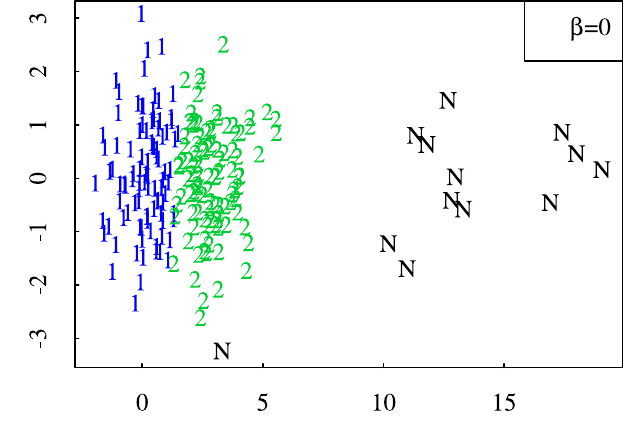}
  \hfill
  \includegraphics[width=0.49\textwidth]{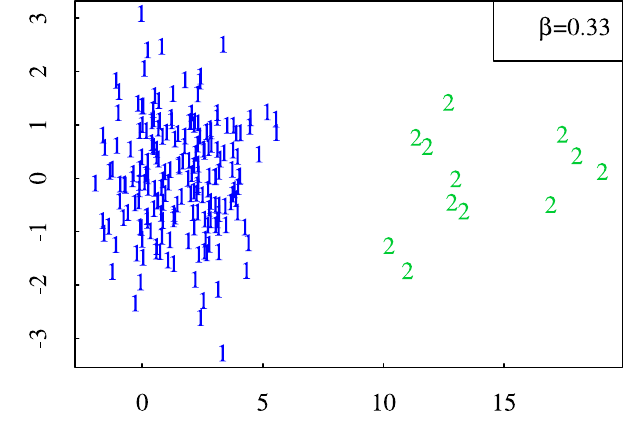}
  \caption{Clusterings with $\beta=0$ and $\beta=\frac{1}{3}$ on artificial data example.}
  \label{fig_beta_data}
\end{figure}

Figure \ref{fig_beta_data} shows a situation in which the choice of $\beta$ affects the clustering a lot. The dataset consists of 100 observations each of ${\cal N}(0,1)$ and ${\cal N}(3,1)$ along the $x$-axis and 12 observations from ${\cal N}(12,25)$. OTRIMLE was fitted with $G=2$. The first two mixture components are not very well separated. $\beta=0$ does not penalize noise, and therefore the observations from the third component are declared ``noise'' and the first two components are separated. However, $\beta=\frac{1}{3}$ merges the first two components and declares the third one the second cluster. The ``switching point'' between these two ways of ``interpreting'' the clustering structure is at about $\beta=0.3$; larger values of $\beta$ don't change the clustering anymore. Despite the ``true'' $G$ being 3 here, regarding interpretation, depending on the application it may well make sense to either treat the smallest mixture component as noise/outliers, or to merge the first two components to a single cluster. $\beta$ tunes the method to rather produce noise, or to rather tolerate not-so-Gaussian components in cases like this.

\section{Applications}  \label{sec_application}
In this Section we apply OTRIMLE and the alternative methods mentioned before
to two real datasets. The first example does not come with any ``ground 
truth'', whereas the second one has ``true'' classes. 
\subsection{Dortmund data} \label{sec_dortmund} 
We here analyze a data set giving information about 170 districts of the 
German city of Dortmund, which is described in \cite{Sommerer_Weihs_2005}. 
We used five sociological key variables 
and transformed them in such a way that fitting Gaussian distributions 
within clusters makes sense. The resulting variables are the logarithm of the unemployment rate (``unemployment''), the birth/death balance divided by number of inhabitants (``birth.death''), the migration balance divided by number of inhabitants (``moves.in.out''), the logarithm of the rate of employees paying social insurance (``soc.ins.emp''), and the square root of the number of inhabitants
(``inhabitants'').

\begin{figure}[!t]
  \centering
  \subfigure[MCLUST without  noise component ]{
    \includegraphics[width=.475\textwidth]{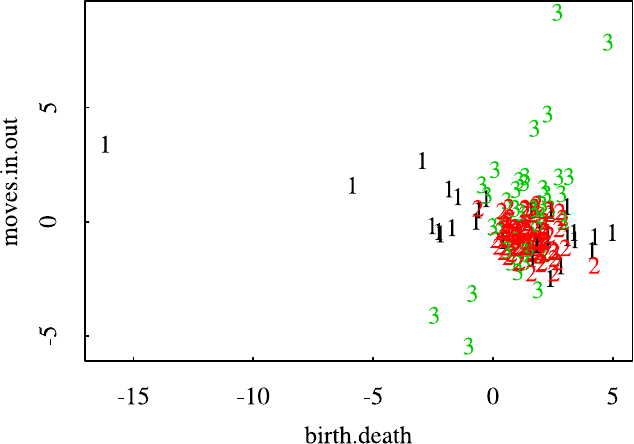}
    \label{fig_dortmund_noiseless}
  }
  \subfigure[OTRIMLE]{
    \includegraphics[width=.475\textwidth]{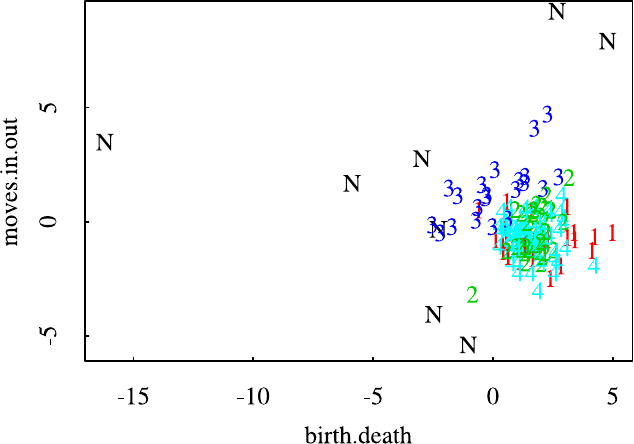}
    \label{fig_dortmund_orimle_1}
  }
   \caption{Scatterplot of {\tt birth.death} and {\tt moves.in.out}
from Dortmund dataset with MCLUST clustering (left)
and OTRIMLE clustering (right) with $G=4$.}
    \label{fig_dortmund_noiseless_orimle}
\end{figure}
\begin{figure}[!t]
  \centering
  \subfigure[OTRIMLE]{
    \includegraphics[width=.475\textwidth]{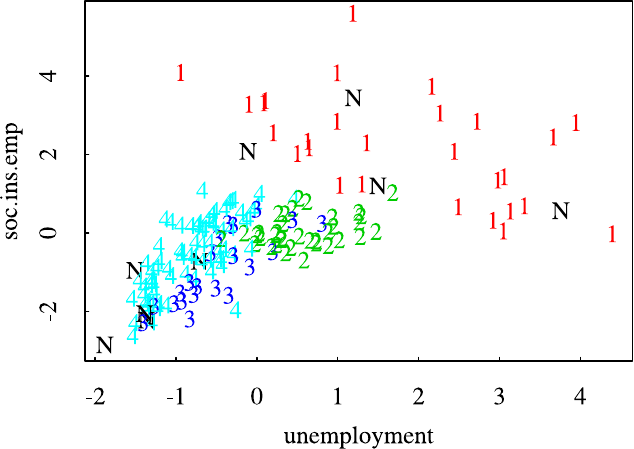} 
    \label{fig_dortmund_orimle_2}
  }
  \subfigure[TCLUST with 7\% trimming]{
    \includegraphics[width=.475\textwidth]{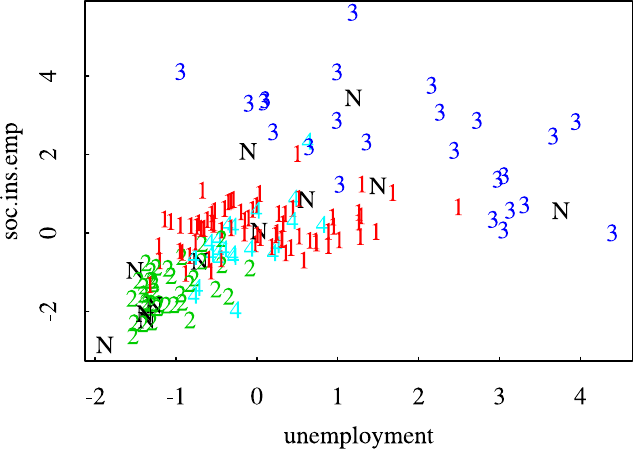} 
    \label{fig_dortmund_tclust}
  }
    \caption{Scatterplot of {\tt soc.ins.emp} and {\tt unemployment}
from Dortmund dataset with OTRIMLE clustering (left)
and TCLUST clustering with trimming rate 7\% (right) with $G=4$.}
  \label{fig_dortmund_orimle}
\end{figure}
All variables were centered and standardised by the median absolute deviation.
Figure \ref{fig_dortmund_noiseless_orimle} shows a scatterplot of 
{\tt birth.death} and {\tt moves.in.out}. 
In order to deal with
overplotting, an existing extreme outlier with values 
$\approx (-200,50)$ is not shown. 
Figure \ref{fig_dortmund_orimle}
shows a scatterplot of 
{\tt unemployment} and {\tt soc.ins.emp}. 
Figure \ref{fig_dortmund_orimle} shows some more moderate outliers. 
The left side of
Figure \ref{fig_dortmund_noiseless_orimle} shows a clustering from 
fitting a plain Gaussian mixture with $G=4$ to all five variables by 
R's MCLUST package. Cluster 4 is a one-point
cluster made of the extreme outlier. Clusters 1 and 3 basically
fit two different varieties of moderate outliers, 
whereas all the more than 120 districts that
are not extreme regarding these two variables are put together in a single
cluster. Clearly, it would be more desirable to have a clustering that
is not dominated so much by a few odd districts, given that there is 
some meaningful structure among the other districts. Such a clustering
is produced by the OTRIMLE method, shown on the right side of Figure 
\ref{fig_dortmund_noiseless_orimle} and on the left side of 
Figure \ref{fig_dortmund_orimle}. The clustering is nicely interpretable with
cluster no. 3 collecting a group of districts with higher migration balance and
very scattered birth/death rate, cluster no. 1 being a high variation cluster characterized by high unemployment or high number
of employees paying social insurance, cluster no. 2 being a homogeneous
group with medium number of employees paying social insurance and rather high
but not very high unemployment, and cluster no. 4 collecting most districts
with low values on both of these variables.
For this data set, values between $\beta=0$ and $\beta=0.5$ 
yield the same clustering (with eigenratio constraint $\gamma=20$), 
despite the fact that $\beta=0$ leads to
$\log(\delta)=-11.5$ and $\pi_{0,n}=0.055$, whereas for $\beta=0.5$, \eqref{eq_orimle} produces $\log(\delta)=-11.9$ and $\pi_{0,n}=0.054$.

Looking at the methods introduced in Section \ref{sec_methods_from_literature}, it turns out that substantial disagreement may exist between  different robust clustering methods. Applying the methods implemented in 
{\tt tclust} and discussed in \cite{GarciaEscudero_Gordaliza_2011},  $\alpha=0.07$ was found to be a good trimming rate for TCLUST with $G=4$ here. 
The resulting clustering is compared with OTRIMLE's in Figure \ref{fig_dortmund_orimle}. Although what was trimmed is almost identical to OTRIMLE's ``noise'', the clustering is somewhat different, with TCLUST's clusters no. 1 and 4 ranging into the area of ``high variation characterized by high unemployment or 
high number
of employees paying social insurance'' as mainly represented by cluster no. 3 here and cluster 1 of OTRIMLE. It is hard to interpret OTRIMLE's cluster no. 4 using any pair of variables. This can be seen in  the online supplement (\cite{Coretto_Hennig_2014_supplement}), as well as solutions of the other methods.

\subsection{Folk song data} \label{sec_folk}
The second dataset was provided by Daniel M\"ullensiefen. The observations are 776 folk song melodies, 586 of which are from Luxembourg and the remaining 190 are from Warmia in Poland. These are the ``true'' classes. The melodies are originally from the ESAC melody database (\cite{Schaffrath92}). The 18 features (see the online supplement (\cite{Coretto_Hennig_2014_supplement}) for a list) were computed by the software 
``FANTASTIC'' (\cite{Muellensiefen09}). 

Visual inspection reveals that there are many unusual melodies, i.e., outliers in the dataset. The main bulks of melodies from Luxembourg and Warmia differ systematically from each other, although there is much overlap and no strong separation. 
For measuring to what extent clusterings computed with $G=2$ coincided with the two regions, we used the adjusted Rand index (ARI; \cite{HubertArabie1985}) with an expected value of 0 for two random clusterings and a maximum of 1 for perfect agreement. %
OTRIMLE with settings as above ($\beta=0, \gamma=20$) classified 36.9\% of the observations as ``noise''. %
The ARI between the OTRIMLE solution and the original regions is 0.155. %
For this (as for the other clustering methods), the OTRIMLE solution was interpreted as a three-cluster solution with ``noise'' as third cluster. %
Default MCLUST yields an ARI$=-0.045$, %
MCLUST with noise yields ARI$=-0.017$, %
ot.tclust yields ARI$=0.016$ %
(the original TCLUST function with trimming rate 0.369 as suggested by OTRIMLE above achieves ARI$=0.089$)   and %
tmix yields ARI$=0.083.$ %
OTRIMLE's ARI-value, though clearly better than that of the other methods, is not particularly high, but computing the ARI only on the observations that were not classified as noise achieves ARI$=0.392$ (the solution with $\beta=\frac{1}{3}$ is slightly worse here but slightly better above regarding the ARI including the noise points), which suggests that there is a clear correspondence between OTRIMLE's clustering and melodies that are typical for the regions.


\section{Concluding Remarks}\label{sec_concluding_remarks}

Despite our effort to make the simulation study fair, ultimately it would be good to have comparisions of methods run by researchers who did not have their hand in the design of any of the methods.
Every method was best for certain DGPs in the simulation study, and simulation studies could be designed that make any method ``win''. Readers need to make up their own mind about to what extent our study covered situations that are important to them. One of our major aims was to confront all methods with DGPs that do not exactly match their model assumptions, but for which the methods nevertheless could be legitimately used.  
In fact, we incorporated crucial ideas from both MCLUST (initialization) and TCLUST (eigenvalue ratio constraints), and the combination of these ideas used here could actually be beneficial for all methods.  The methods presented in this paper will soon be available in the new R-package OTRIMLE.

The problem of determining the number of clusters $G$ is very important in  practice. Here are two possible approaches for RIMLE.  Firstly, the most popular approach for fitting plain mixture models, namely the  Bayesian Information Criterion, can be used (treating RIMLE/OTRIMLE improper constant density as a proper one),  \cite{Fraley_Raftery_1998}. Secondly, $G$ could be decided in an exploratory way by monitoring the changes of the pseudo-likelihood over different values of $G$ in a similar way to what is done for TCLUST in \cite{GarciaEscudero_Gordaliza_2011}. 
Investigating these approaches in depth is beyond the scope of this paper.


\bibliographystyle{chicago}
\bibliography{references}%
\end{document}